\documentclass[eqsecnum,showpacs,twocolumn]{revtex4}
\usepackage{xspace,amsfonts,amssymb,graphicx}
\begin{document}
\title{Directly Observing Momentum Transfer in Twin-Slit ``Which-Way''
Experiments}
\author{H. M. Wiseman}
\email{H.Wiseman@griffith.edu.au} 
\affiliation{Centre for Quantum Dynamics, School of Science, Griffith
University,  Brisbane, 4111 Australia. \\
Phone: +61 7 3875 7271  \ \  
Facsimile: +61 7 3875 7656}

\newcommand{\beq}{\begin{equation}}
\newcommand{\eeq}{\end{equation}}
\newcommand{\bqa}{\begin{eqnarray}}
\newcommand{\eqa}{\end{eqnarray}}
\newcommand{\nn}{\nonumber}
\newcommand{\nl}[1]{\nn \\ && {#1}\,}
\newcommand{\erf}[1]{Eq.~(\ref{#1})}
\newcommand{\erfs}[2]{Eqs.~(\ref{#1})--(\ref{#2})}
\newcommand{\dg}{^\dagger}
\newcommand{\rt}[1]{\sqrt{#1}\,}
\newcommand{\smallfrac}[2]{\mbox{$\frac{#1}{#2}$}}
\newcommand{\half}{\smallfrac{1}{2}}
\newcommand{\bra}[1]{\langle{#1}|}
\newcommand{\ket}[1]{|{#1}\rangle}
\newcommand{\ip}[2]{\langle{#1}|{#2}\rangle}
\newcommand{\sch}{Schr\"odinger }
\newcommand{\schs}{Schr\"odinger's }
\newcommand{\hei}{Heisenberg }
\newcommand{\heis}{Heisenberg's }
\newcommand{\bl}{{\bigl(}}
\newcommand{\br}{{\bigr)}}
\newcommand{\ito}{It\^o }
\newcommand{\str}{Stratonovich }
\newcommand{\dbd}[1]{{\partial}/{\partial {#1}}}
\newcommand{\sq}[1]{\left[ {#1} \right]}
\newcommand{\cu}[1]{\left\{ {#1} \right\}}
\newcommand{\ro}[1]{\left( {#1} \right)}
\newcommand{\an}[1]{\left\langle{#1}\right\rangle}
\newcommand{\st}[1]{\left|{#1}\right|}
\newcommand{\implies}{\Longrightarrow}
\newcommand{\tr}[1]{{\rm Tr}\sq{ {#1} }}
\newcommand{\del}{\nabla}
\newcommand{\du}{\partial}
\newcommand{\tick}{$\sqrt{\phantom{I_{I}\hspace{-2.1ex}}}$}
\newcommand{\cross}{$\times$}
\newcommand{\ww}{which-way }
\newcommand{\ps}[1]{\hspace{-5ex}{\phantom{\an{X_{w}}}}_{#1}\!}

\begin{abstract}
Is the destruction of interference by a which-way measurement 
due to a random momentum transfer $\wp\agt\hbar/s$, with $s$ 
the slit separation? The weak-valued probability distribution 
$P_{\rm wv}(\wp)$, which is {\em directly observable}, 
provides a subtle answer. 
$P_{\rm wv}(\wp)$ cannot have support on the
interval $[-\hbar/s,\hbar/s]$. Nevertheless, its moments can be 
identically zero. 

\end{abstract}


\maketitle

\section{Introduction}
Making a position measurement to determine through which slit
 a particle has passed necessarily destroys the twin-slit interference
pattern (see Fig.~1).  This is the canonical example of Bohr's
complementarity principle
\cite{BohrEinst,WheZur83}. To defend this principle against Einstein's
recoiling slit {\em gedankenexperiment},
Bohr relied upon the recently (in 1927)
derived Heisenberg uncertainty relation \cite{Hei27} to show that
the position measurement would cause an ``uncontrollable change in
the momentum'' $\wp \agt \hbar/s$, where $s$ is the slit separation
\cite{BohrEinst}.
This is just what is required to wash out the interference
pattern, thereby enforcing complementarity. Bohr's argument was
famously reiterated by Feynman \cite{FeyLeiSan65}, for a measurement
using \heis light microscope \cite{Hei27}.

\begin{figure}[tbp]
\vspace{-0.5ex}
\includegraphics[angle=90,width=0.45\textwidth]{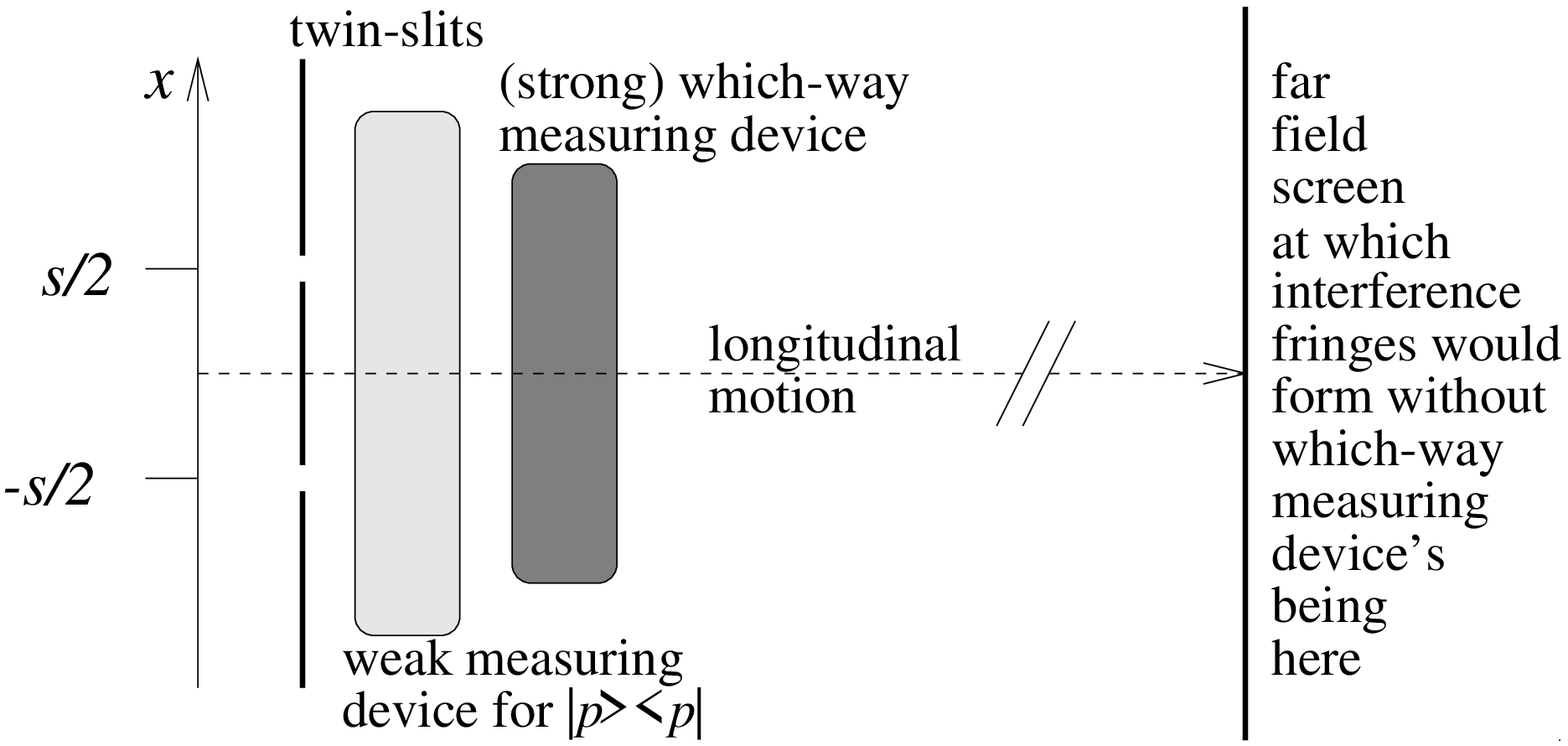}
\vspace{0.5ex}
   \caption{Diagram	of a twin-slit \ww experiment. The 
   initial state $\ket{\psi_{i}}$ is formed by the slits and 
   propagates longitudinally towards the final screen. This  is 
   in the far field, so detecting the position of 
the particle there is equivalent to measuring its final momentum 
$p_{f}$. With the which-way measurement (WWM) device in place 
the distribution for 
$p_{f}$ is just the single-slit diffraction pattern. In its absence, 
a twin-slit interference pattern would form. 
Between the slits and the WWM device is a device which makes a 
weak measurement of $\ket{p_{i}}\bra{p_{i}}$, as 
explained later	in the text.}
   \label{fig1}
\end{figure}

In 1991 Scully, Englert and Walther \cite{ScuEngWal91}
proposed a new which-way ({\em welcher Weg}) scheme for which, they
calculated, no momentum would be transferred to the particle. Thus, they
concluded, the arguments of Bohr and Feynman
were wrong in general, and that complementarity must be
 deeper than uncertainty. Storey, Tan, Collett and
Walls (STCW)  \cite{StoTanColWal94} disagreed, proving a theorem that
if interference is destroyed then there is some transverse 
momentum transfer
$\geq \hbar/s$. The debate \cite{QI94,Nature95} was partially resolved
when it was pointed out \cite{WisHar95} that SEW and STCW were using different
concepts of ``momentum transfer'', and that their analyses were in
fact complementary. Aspects of both are found
in two further characterizations of momentum transfer that
have since been proposed \cite{Wis97a,Wis98a}. However, these
characterizations are not
obtainable from experiments directly (i.e.
in a way that would make sense to a classical physicist).

In this Letter I propose a new and attractive resolution: a way to
directly observe a single probability distribution $P(\wp)$
for momentum transfer $\wp$ in a
\ww measurement (WWM). My method shows that SEW were right,
in that the variance of $P(\wp)$ is zero for their scheme,
but that STCW were also
right, in that $P(\wp)$ cannot have support on the interval 
$[-\hbar/s,\hbar/s]$.
This is possible only because $P(\wp)$ must be measured as
a {\em weak valued} \cite{AhaAlbVai88} probability distribution, 
and hence may
take negative values.

The body of this Letter is organized as follows. First I
introduce the general formalism of WWMs, and use this to
explain the four characterizations of momentum transfer proposed in
Refs.~\cite{ScuEngWal91,StoTanColWal94,Wis97a,Wis98a}.
I discuss these characterizations in terms of 6
key properties, showing that each lacks at least 2 properties. This
motivates the introduction of the present approach, based on weak
values, which has all properties except non-negativity of $P(\wp)$.

\section{Which-way measurements}

Say the two slits in the Young's interferometer 
are centred at $x=\pm s/2$. Note that it is 
necessary to restrict the discussion to an interferometer of this 
kind, where the initial superposition is in transverse position and the free 
evolution preserves the conjugate quantity (transverse momentum), so 
that the issue of loss of visibility related directly to 
transverse momentum transfer. Analogous issues may be identifiable for 
other sorts of interferometer, but that is beyond the scope of this 
Letter. 
 
A generalized measurement of $x$ transforms the
initial state produced by the twin slits, $\ket{\psi_{i}}$, into a final mixed
state, $\rho_{f}$, according to \cite{StoTanColWal94,Wis97a}
\beq \label{transform}
\rho_{f} =  \hat{O}_{\xi}\ket{\psi_{i}}\bra{\psi_{i}}\hat{O}_{\xi}\dg.
\eeq
Here $\cu{\xi}$ is the set of measurement results, which here is summed 
over, as I am using the {\em Einstein summation convention}.  This gives 
the final state {\em averaged} over all possible results, 
because  $\hat{O}_{\xi}\ket{\psi_{i}}$ represents 
the unnormalized state given the result $\xi$, and its  modulus squared 
is the probability to obtain that result $\xi$. 

Note that there are infinitely 
many sets of operators $\{\hat{O}_{\xi}\}$ which give the {\em same} 
average transformation (\ref{transform}) \cite{TanWal93}. These different sets 
 can be obtained physically by using  the same measurement 
interaction between the system and the apparatus, but from 
resolving this apparatus in different bases. 
For example, for many measurements there is a ``quantum
eraser'' basis \cite{ScuDru82} complementary to the basis that gives 
the best \ww information. In these different bases, the  
states  $\hat{O}_\xi \ket{\psi_{i}}$ conditioned on individual results may be 
completely different, but the average state $\rho_f$ is the same. 

For \erf{transform} to describe a position measurement, it is 
necessary to restrict the operators $\hat O_\xi$ to be functions of the 
position operator $\hat{x}$. Then in the position representation one 
finds
\beq \label{multpos}
\bra{x}\hat{O}_{\xi}\ket{\psi_{i}} = O_{\xi}(x) \psi_{i}(x).
\eeq
Thus the measurement is defined by
 the functions $\cu{O_{\xi}(x)}$, which
 are constrained mathematically only by the completeness condition
\beq \label{complet}
\forall x\;   O_{\xi}(x)O_{\xi}^{*}(x) = 1 
\eeq
which ensures that $\rho_{f}$ is normalized. 
For narrow slits, [i.e. $|\psi(x)|^{2} \simeq
\delta(2x+s)+\delta(2x-s)$],
the visibility of the far field interference pattern
can be shown
\cite{Wis97a} to be given by
\beq \label{vis}
V = \st{  O_{\xi}(-s/2)O_{\xi}^{*}(s/2)}.
\eeq

It is evident from \erf{multpos} that if all of the $O_{\xi}(x)$ are
flat in the region of the slits, then a {\em single slit} initial
wavefunction $\psi_{i}(x)$ will emerge unchanged from the measurement
region. In particular, its momentum distribution (the diffraction
pattern) will be no wider than
that with no measurement. This phenomenon is compatible with
the complete loss of
visibility (\ref{vis}), and indeed would occur for the 
micromaser experiment
proposed by SEW. It is by this directly observable measure
that SEW claimed there would be no momentum transfer.

In response,
STCW proved that Eqs.~(\ref{complet}) and (\ref{vis}) imply that
for at least one $\xi$, the Fourier transform
\beq \label{fourier}
\tilde{O}_{\xi}(p) = (2\pi \hbar)^{-1/2}\int\! dx\, O_{\xi}(x) e^{-ix p/\hbar}
\eeq
does not have support on the interval $(-\hbar/s,\hbar/s)$ if $V=0$. The
significance of this is that in the momentum representation
\erf{multpos} becomes
\beq \label{convmom}
\bra{p}\hat{O}_{\xi}\ket{\psi_{i}}
 = \int\! d\wp\, \tilde{O}_{\xi}(\wp) \tilde\psi_{i}(p-\wp).
\eeq
This implies that $\tilde{O}_{\xi}(\wp)$ can be interpreted as a
momentum transfer {\em amplitude} distribution. Moreover, from
\erf{convmom}, if the
initial state $\ket{\psi_{i}}$ were a $p=0$ momentum eigenstate then
 the final momentum distribution would be
$\bra{p}\rho_{f}\ket{p} =  \tilde{O}_{\xi}(p)\tilde{O}_{\xi}^{*}(p)$. 
Thus the
momentum transfer in the STCW theorem is also directly observable, as 
the creation of momentum components greater than or equal to 
$\hbar/s$.

For the Einstein
recoiling slit \cite{BohrEinst} and
Feynman light microscope \cite{FeyLeiSan65}, it turns out that
\beq \label{classmt}
\bra{p}\rho_{f}\ket{p} = \int\! d\wp\, P_{\rm classical}(\wp)
|\tilde{\psi}_{i}(p-\wp)|^{2}
\eeq
where $P_{\rm classical}(\wp)$ is non-negative. For such schemes the 
$O_{\xi}(x)$ can be chosen as $\sqrt{N_{\xi}}\exp(-ik_{\xi}x)$, and $P_{\rm 
classical}(\wp) = \sum_{\xi} N_{\xi}\delta(\wp-\hbar k_{\xi})$. 
For these schemes, the final momentum distribution
is broadened identically for all initial states, and the SEW and
STCW characterizations of momentum transfer agree.
However, for some schemes (most notably that of SEW)
the functions $O_{\xi}(x)$ are necessarily not of this form,
 which is why more than one characterization of momentum transfer is
possible in general. These facts were first pointed out
by Wiseman and Harrison \cite{WisHar95}.

The essential problem is that if the momentum transfer
(\ref{convmom}) is not of classical form (\ref{classmt}), 
then the transfer of
momentum is not clearly defined for an initial twin-slit state, as it
does not have definite momentum. The obvious solution is to use a
quantum formalism in which a momentum is associated with 
the particle even when it is not in a momentum eigenstate. 
Two such formalisms have been
investigated in the past.

In the Wigner function formalism
\cite{Wis97a}, the particle is described by a pseudo probability
distribution on phase space $W(x,p)$ which gives the correct
 marginal  distributions for $x$ and $p$.
The transformation \erf{transform} becomes
\beq
W_{f}(p,x) =
\int\!  d\wp\, W_{i}(x,p-\wp) P_{\rm Wigner}(\wp;x).
\eeq
Here $P_{\rm Wigner}(\wp;x)$ is defined in terms of $\cu{O_{\xi}}$ and acts
formally as an $x$-dependent momentum transfer probability distribution.
Both the local momentum transfer at $x=\pm s/2$, and
the nonlocal momentum transfer at $x=0$ (midway between the slits) are
relevant to the final momentum distribution.

In the Bohmian formalism \cite{Wis98a}, an individual particle 
has a definite
position $x$ and momentum $p=m\dot{x}
= {\rm Re}[-i\hbar\psi'(x)/\psi(x)]$.
The probability distribution for $x$ is
the usual $|\psi(x)|^{2}$, but that for $p$ equals
$|\tilde{\psi}(p)|^{2}$  only in the far field  \cite{Boh52}. 
By following the
trajectories of individual particles one
can calculate a time-dependent momentum transfer probability distribution
$P_{\rm Bohm}(\wp;t)$ where $t$ is the time after the WWM \cite{Wis98a}.
In this formalism $P_{\rm Bohm}(\wp,0^{+})$ gives
the local momentum transfer, but the momentum continues to change
after the measurement (i.e. nonlocally), until the far field
($t\to\infty$).

\section{Properties of $\wp$-characterizations}
By considering the above four characterizations it is possible to
identify certain key desirable properties. They are desirable in the
sense that classically they would be properties of a complete
description of momentum transfer. They are key in the sense that all
are properties of some of the above characterizations, but none are
properties of all. The details are given in table~1.

\begin{table*}
    \caption{Table of properties of various proposed methods of
    characterizing the momentum transfer in a WWM.
    A $\sqrt{}$ or
    $\times$ indicates the presence or absence
    of a property, and a $-$ indicates non-applicability.}
	\label{table:properties}
	\centering
	\begin{ruledtabular}
	\begin{tabular}{|l|c|c|c|c|c|c|c|}
	\hline
	Characteriz- & Ref. & Is it described & Is
	$P(\wp)$
	& Does it reflect & Does it
	reflect & Is it directly & Is it independent \\
	ation of $\wp$ &  & by a $P(\wp)$?
	&  positive? & $\Delta P(p)$? & $\Delta \an{p^{n}}$? & observable? &
	of the basis $\{\xi\}$?  \\
	\hline\hline
	SEW & \cite{ScuEngWal91} & \cross & $-$ & \cross & \tick &
	\tick & \tick  \\
	\hline
	STCW & \cite{StoTanColWal94} & \cross & $-$ & \tick & \cross &
	\tick & \tick  \\
	\hline
	$P_{\rm Wigner}(\wp;x)$ & \cite{Wis97a} & \tick & \cross & \tick &
	\tick & \cross & \tick  \\
	\hline
	$P_{\rm Bohm}(\wp;t)$ & \cite{Wis98a} & \tick & \tick & \tick &
	\tick & \cross & \cross  \\
	\hline
	$P_{\rm wv}(\wp)$ & here &  \tick & \cross & \tick & \tick & \tick
& \tick
	\\
	\hline
\end{tabular}
\end{ruledtabular}
\end{table*}

The first property is that there should be a probability distribution
$P(\wp)$ in order to fully characterize $\wp$.
This is as opposed to partial
characterizations, such as
the (lack of) increase in the width of the single-slit momentum
distribution (SEW), or the creation of momentum values outside a certain
range from a $p=0$ momentum eigenstate (STCW). 
The second property is that $P(\wp)$,
if it exists, should be non-negative.

The third property is that $\wp$ should reflect any change in
the momentum distribution. Clearly in a \ww scheme the momentum
distribution does change (the fringes disappear) so if $\wp$ is
characterized by a distribution, it should not equal $\delta(\wp)$.
The fourth property is that $\wp$ should reflect the change in the
{\em moments} of the momentum. If the single-slit diffraction patterns are
unaffected by the WWM, as in the scheme of SEW, then it
turns out \cite{Wis97a} that the moments of the
twin-slit interference pattern are also {\em exactly} unchanged.
For such schemes, the moments of $\wp$ should be zero also.

The fifth property is that the characterization of $\wp$ should be
directly observable, in the sense defined in the introduction.
In particular, no knowledge of quantum mechanics (such as the wave
properties of particles, or even the value of $\hbar$) should
be required for understanding this part of the experiment, or for
processing the data.

The sixth and final property is that the characterization should not
depend upon the basis into which the measurement apparatus is resolved.
That is, for a given measurement interaction, giving the 
transformation (\ref{transform}), the characterization 
 should not depend upon the particular set of functions 
$\{O_{\xi}\}$ arising from resolving the apparatus in a particular 
basis (as discussed in Sec.~II). For example,  
the momentum transfer should be the same in the quantum
eraser basis \cite{ScuDru82} as in the \ww basis.

\section{Weak Values} The proposal in this Letter is to characterize $\wp$
using the theory
of weak values \cite{AhaAlbVai88}. A weak value is a the mean value
of the result of a weak measurement of some quantity $\hat{A}$. This
is a measurement which yields an arbitrarily small amount of information
about $\hat{A}$, and disturbs the system correspondingly weakly in doing
so. This means that the ensemble giving the average must be
correspondingly larger than that which would be needed for averaging
a strong (projective) measurement.

Weak values are interesting (for example, lying outside the range
of eigenvalues of $\hat{A}$ \cite{AhaAlbVai88})
in the case when they are post-selected
on the obtaining of a certain result from a
projective measurement at a later time.
If the initial state is $\ket{\psi_{i}}$ and the projector for the
desired final measurement result is $\ket{\phi_{f}}\bra{\phi_{f}}$, then the
weak value of $\hat{A}$ conditioned on this final result is
\cite{AhaAlbVai88}
\beq
\ps{\bra{\phi_{f}}}\an{{A}_{w}}_{\ket{\psi_{i}}} = {\rm
Re}\frac{\bra{\phi_{f}}\hat{A}\ket{\psi_{i}}}{\ip{\phi_{f}}{\psi_{i}}}.
\eeq
This prediction was soon verified experimentally \cite{RitStoHul91}.
Since then post-selected weak values have found many
applications, including defining tunneling time in a directly
observable manner \cite{Ste95}, and resolving quantum paradoxes
\cite{Aha01}. With a few simple generalizations, weak values have
also been found \cite{Wis02a} to explain puzzling aspects of a
well-known cavity QED experiment \cite{FosOroCasCar00}.

Given these successes of weak value theory, it is natural to apply
it to the question of momentum transfer in WWMs.
A schematic experiment is shown in Fig.~1. After the
initial state $\ket{\psi_{i}}$ is formed, but before the WWM,
a weak measurement of the projector
$\ket{p_{i}}\bra{p_{i}}$ is made. If this were a strong measurement
then the mean value (whether post-selected or not)
would be $|\ip{p_{i}}{\psi_{i}}|^{2}$, the probability
for the initial momentum to be $p_{i}$.  For a weak measurement, an
individual result is given by
$|\ip{p_{i}}{\psi_{i}}|^{2} + \sigma S$, where $S$ is a standard
normal variable and $\sigma^{2}$ is an arbitrarily large parameter (equal
to $1/dt$ in Ref.~\cite{Wis02a}).
The quantum back-action of this measurement on the
system state is, to leading order in $\sigma^{-1}$,
\beq
\ket{\psi_{i}} \to
\sq{1  + \ro{ \ket{p_{i}}\bra{p_{i}}
-|\ip{p_{i}}{\psi_{i}}|^{2}}S/2\sigma }\ket{\psi_{i}}.
\eeq

Following the weak measurement, there is
the WWM represented by the operators $\{\hat{O}_{\xi}\}$ in
\erf{transform}. The post-selecting measurement is a strong
measurement of the final momentum $p_{f}$.  The weak measurement
and final measurement can be constructed as appropriate for a
classical particle. Thus, the post-selected weak
value would classically be interpreted as $P(p_{i}|p_{f})$. Since free
evolution preserves momentum, if there were no WWM then one
would expect that $P(p_{i}|p_{f}) = \delta(p_{i}-p_{f})$,
and any deviation from this due to the WWM would be
interpreted as a momentum transfer.

Using the generalized theory of Ref.~\cite{Wis02a},
the {\em weak valued} conditional probability $P_{\rm wv}(p_{i}|p_{f})$  
which would be measured for a quantum system is
\beq
\ps{\bra{p_{f}}}
\an{\ket{p_{i}}\bra{p_{i}}}_{\ket{\psi_{i}}}
= {\rm Re}\frac{\bra{p_{f}}\hat{O}_{\xi}
\ket{p_{i}}\ip{p_{i}}{\psi_{i}}\bra{\psi_{i}}\hat{O}_{\xi}\dg\ket{p_{f}}}
{\bra{p_{f}}\hat{O}_{\zeta}\ket{\psi_{i}}
\bra{\psi_{i}}\hat{O}_{\zeta}\dg\ket{p_{f}}}. \label{condtal}
\eeq
(Recall the summation convention.) Now the joint and conditional
distributions are related by $P(p_{i},p_{f})=P(p_{i}|p_{f})P(p_{f})$.
Since the weak measurement introduces negligible
disturbance to the system, the denominator in \erf{condtal} is simply
$P(p_{f})$, so
\beq
P_{\rm wv}(p_{i},p_{f}) =   {\rm Re}\cu{\bra{p_{f}}\hat{O}_{\xi}
\ket{p_{i}}\ip{p_{i}}{\psi_{i}}\bra{\psi_{i}}\hat{O}_{\xi}\dg\ket{p_{f}}}.
\eeq
We can rewrite this as a function $P_{\rm wv}(\wp,p_{i})$ of
the momentum transfer $\wp = p_{f}-p_{i}$.
If one then repeated the experiment for all values of $p_{i}$, one
could integrate this over all $p_{i}$ to finally
obtain the weak-valued probability
distribution as
\bqa
 P_{\rm wv}(\wp) &=& \int\! dp_{i}\, {\rm Re} 
\left\{ \bra{p_{i}+\wp}\hat{O}_{\xi}
\ket{p_{i}} \right. \nn \\ && \times \left.   \ip{p_{i}}{\psi_{i}}\bra{\psi_{i}}\hat{O}_{\xi}\dg\ket{p_{i}+\wp}\right\}.  
\label{wvmt1}
\eqa

It is now straightforward to evaluate \erf{wvmt1} using
\erf{transform}. The resulting distribution for
the momentum transfer caused by the
WWM is remarkably elegant:
\beq \label{wvmt2}
P_{\rm wv}(\wp) = {\rm Re}\cu{ \tilde{O}_{\xi}(\wp)
\tilde{Q}_{\xi}^{*}(\wp)}.
\eeq
The Fourier transform is as defined in \erf{fourier}, and
\beq
Q_{\xi}(x) = O_{\xi}(x) |\psi_{i}(x)|^{2}.
\eeq
Thus the probability distribution for $\wp$ depends upon the initial
state in a very natural way. Note that $P_{\rm wv}(\wp)$  may be
negative, although for classical momentum
kicks $P_{\rm wv}(\wp) = P_{\rm
classical}(\wp) \geq 0$. If there is no WWM,
so that $O_{\xi}(x) \equiv 1$, then $P_{\rm
wv}(\wp)=\delta(\wp)$ as one would wish.

\section{Properties of $P_{\rm wv}(\wp)$}

From \erf{wvmt2} and \erf{complet}
a number of results are easily proven using the
moment-generating function
\bqa
\Phi(q) &=& \int \!d\wp\, P_{\rm wv}(\wp) e^{i\wp q/\hbar} \\
&=&  {\rm Re} \int \!dx\, |\psi_{i}(x)|^{2}O_{\xi}(x)O_{\xi}^{*}(x-q).
\label{mgf}
\eqa
First, $P_{\rm wv}(\wp)$ is normalized, as $\Phi(0)=1$. Second, using
the Schwartz inequality, $|\Phi(q)|\leq 1$, as it would be for a
true (non-negative) probability distribution.
Third, since the moments of $\wp$ are
given by
\beq
\an{\wp^{n}}_{\rm wv} = \ro{-i\hbar \dbd{q}}^{n}\Phi(q)|_{q=0},
\eeq
it follows from \erf{mgf} that if the $O_{\xi}$ are flat (i.e. have
all derivatives zero) in the region
of the slits where $|\psi_{i}(x)|^{2}$ is nonzero, then all of the moments
of $\wp$ are zero. Thus the claim of SEW, that their scheme would not
transfer any momentum to the particle, could be validated 
experimentally by calculating the moments of $\wp$ 
using the measured $P_{\rm wv}(\wp)$.

Fourth, despite this last fact, $P_{\rm wv}(\wp)$ also reflects the change
in the
momentum distribution caused by a WWM. For narrow slits at $x=\pm s/2$,
\beq
\Phi(s) =  \frac12 {\rm Re} \sq{O_{\xi}\ro{-3\frac{s}{2}}O_{\xi}^{*}\ro{-\frac{s}{2}} 
+ O_{\xi}\ro{-\frac{s}{2}}O_{\xi}^{*}\ro{\frac{s}{2}}}.
\eeq
Using \erf{complet} and \erf{vis} and the triangle inequality, it can 
be proven 
that $|\Phi(s)| \leq 1/2$ for $V=0$. Since in addition $\Phi(0) = 1$ and
$|\Phi(q)| \leq 1\;\forall q$, it follows from the theorem in Appendix A of
Ref.~\cite{Wis97a} that
\beq
{\rm Support}[P_{\rm wv}(\wp)] \not\subset 
(-\pi\hbar/3s,\pi\hbar/3s). \label{support}
\eeq
This establishes the weaker result, quoted in the abstract and  Sec.~I,  that the support
of $P_{\rm wv}(\wp)$ is not contained in $[-\hbar/s,\hbar/s]$ 
\cite{fn1}.

To illustrate the above results, consider the minimal WWM: a projective
measurement of the sign of $x$, so that $O_{\pm}(x) = \Theta(\pm x)$.
For narrow slits, \erf{wvmt2} gives
\beq \label{singular}
P_{\rm wv}(\wp) = \frac{1}{2}\left[ \delta(\wp) + \frac{\sin \wp
s/2}{\pi \wp} \right].
\eeq
This clearly  satisfies \erf{support}, and yet
has all moments equal to zero  in a distributional sense \cite{Wis97a}. For
example, if the
mathematical approximation of Gaussian slits of width $a$ is used, then
one obtains a distribution whose moments are
well-defined in ordinary calculus, and which vanish exponentially quickly
in $s/a$. 

As noted above, the  property of having zero moments but not being 
a $\delta$-function is possible only because of the fact 
that $P_{\rm wv}(\wp)$ takes negative values. There is no logical 
contradiction here; $P_{\rm wv}(\wp)$ is {\em not} the probability 
distribution for a measurement result. Rather, it is itself the average of 
(weak) measurement results. In classical mechanics, the $P_{\rm wv}(\wp)$ 
so derived would be guaranteed to be the probability distribution for 
the momentum transfer $\wp$, as could be verified  by repeatedly 
measuring $p_{f}$ 
and $p_{i}$, and subtracting them. In quantum mechanics a  
measurement of $p_{i}$ would drastically disrupt the system, so we must be 
content with the first way of finding $P_{\rm wv}(\wp)$, via weak 
values. Negative weak-valued 
probabilities, which have been previously encountered in 
Ref.~\cite{Aha01}, are non-classical, but not nonsensical.

In conclusion, I have shown that using weak measurements one can
directly observe a (weak-valued) probability distribution $P(\wp)$ for
momentum transfer in \ww
measurements. It has all the properties
one would desire of such a distribution except that it may take
negative values. This is unavoidable since for the SEW scheme
the distribution reflects both the fact that all of the moments of
$\wp$ are zero, and the fact that $P(\wp) \neq 0$ for some $\wp$ outside the
interval $[-\hbar/s,\hbar/s]$. Techniques similar to those used in
Ref.~\cite{RitStoHul91} should enable interesting instances of
this distribution, such as that in \erf{singular}, 
to be measured with current technology. 

\begin{acknowledgments}
I would like to thank J. Garretson for a careful reading of this 
work, which  was supported by the Australian Research Council. 
\end{acknowledgments}


\begin{references}
\bibitem{fn1}
{It seems likely that there is a stronger bound using $\pi\hbar/2s$, but
this is as yet unproven.}
\end{references}
\end{document}